\documentclass[12pt,british,english,aip,pof,review]{revtex4-1}
\usepackage[T1]{fontenc}
\usepackage[latin9]{inputenc}
\usepackage{geometry}
\usepackage{amsmath}
\usepackage{amssymb}
\usepackage{stmaryrd}
\usepackage{graphicx}



\usepackage{babel}
\begin{document}
\title{Lock-exchange problem for Boussinesq fluids revisited:\\
Exact shallow-water solution}
\author{Gerasimos Politis}
\author{J\={a}nis Priede}
\altaffiliation[Also at]{ Department of Physics, University of Latvia, Riga, LV-1004, Latvia}

\affiliation{Fluid and Complex Systems Research Centre, Coventry University, Coventry,
CV1 5FB, United Kingdom}
\email{j.priede@coventry.ac.uk}

\begin{abstract}
An exact solution to the lock-exchange problem, which is a two-layer
analogue of the classical dam-break problem, is obtained in the shallow-water
(SW) approximation for two immiscible fluids with slightly different
densities. The problem is solved by the method of characteristics
using analytic expressions for the Riemann invariants. The obtained
solution, which represents an inviscid approximation to the high-Reynolds-number
limit, is in general discontinuous containing up to three hydraulic
jumps which are due to either multivaluedness or instability of the
continuous SW solution. Hydraulic jumps are resolved by applying the
Rankine-Hugoniot conditions for the SW mass and generalized momentum
conservation equations. The latter contains a free parameter $\alpha$
which defines the relative contribution of each layer to the interfacial
pressure gradient. We consider a solution for $\alpha=0,$ which corresponds
to both layers affecting the interfacial pressure gradient with equal
weight coefficients. This solution is compared with the solutions
resulting from the application of the classical Benjamin's front condition
as well as the circulation conservation condition, which correspond
to $\alpha=-1$ and $\alpha\rightarrow\infty,$ respectively. The
SW solution reproduces all principal features of 2D numerical solution
for viscous fluids. The gravity current speed is found to agree well
with experimental and numerical results when the front acquires the
largest stable height which occurs at $\alpha=\sqrt{5}-2.$  We show
that two-layer SW equations for the mass and generalized momentum
conservation can describe interfacial waves containing hydraulic jumps
in a self-contained way without external closure conditions.
\end{abstract}
\maketitle

\section{Introduction}

Although inertia-dominated fluid flows tend to be very complex, especially
in the presence of a free surface or interface, there are certain
hydrodynamic problems of this type which can be solved analytically.
A well-known example is the classic dam-break problem in which an
instantaneous collapse of the reservoir wall causes a mass of water
to be driven by gravity over a horizontal or sloped ground. \citep{Stoker1958,Johnson1997}
This problem was originally solved by \citet{BarredeSaint-Venant1871}
and then shortly afterwards in a more complete form by \citet{Ritter1892}
using the shallow-water (SW) approximation and the method of characteristics.

There is an analogous problem for a stably-stratified two-layer system,
which is known as the lock-exchange problem, where a heavier fluid
in a horizontal channel is initially separated by a vertical lock
from a lighter fluid on the other side. When the lock gate is rapidly
removed, the difference in hydrostatic pressure causes the heavier
fluid to intrude along the bottom into the lighter fluid which in
turn is forced to flow back at the top. Such flows produced by the
lock exchange, which are commonly referred to as gravity currents,
have extensively been studied experimentally due to their occurrence
in various natural and artificial environments\foreignlanguage{british}{.
\citep{Simpson1979,Klemp1994,Shin2004}}

If the heavier fluid is covered by a much lighter ambient fluid, the
lock-exchange problem reduces to that of the dam-break. Likewise,
the two-layer problem reduces to that of the single-layer when the
bottom layer is much thinner than the upper layer which, in this case,
just modifies the effective value of gravity. \citep{Stoker1958}
Although the two-layer problem becomes mathematically equivalent to
that of the single-layer in this limit, the actual flows have substantial
differences. Namely, the heavy fluid is observed to form a bore, i.e.,
a finite-height front, as it propagates along the bottom into the
lighter fluid whereas the ideal dam-break solution predicts a sharp
front edge. \citep{Abbott1961} This has led to a common assumption
that the SW equations for two-layer system are inherently incomplete
and unable to describe internal bores without external closure relations.
\citep{Fyhn2019} The latter have to be deduced by dimensional arguments
\citep{Abbott1961} or derived using various semi-empirical and approximate
integral models. \citep{Baines1995}

For fluids with nearly equal densities, which can be described using
the Boussinesq approximation, \citep{Long1965} the SW equations for
two-layer system bounded by a rigid lid become identical with the
single-layer equations when both are cast in canonical form using
the Riemann invariants. \citep{Ovsyannikov1979} This fact was used
by \citet{Chumakova2009} to suggest that the lock-exchange problem
for Boussinesq fluids is mathematically equivalent to the single-layer
dam-break problem. \citet{Esler2011} argue that this equivalence
is limited by different physical variables of the two-layer problem
mapping to the same Riemann invariants. This makes the inverse mapping
non-unique and results in the lock-exchange flows which have no dam-break
counterparts. However, it has not been realized so far that the lock-exchange
problem for Boussinesq fluids is solvable analytically using the hydrostatic
SW approximation and the method of characteristics like the dam-break
problem. The present paper aims to fill this gap by providing a comprehensive
analytical solution of this generic hydrodynamic problem.

Various hydraulic-type models \citep{Benjamin1968,Huppert1980,Rottman1983}
and approximate \emph{ad hoc} solutions \citep{Keller1991,Lowe2005}
have been proposed for the lock-exchange problem. One of the most
notable is the model of \citet{Shin2004} where the flow is assumed
to consist of a gravity current which is connected to the lock by
a bore propagating upstream. A good agreement with experimentally
observed height of gravity current is achieved when the momentum and
energy are assumed to be conserved by the system of two jumps rather
than by each jump separately. \citet{Borden2013a} find that this
model fares much worse when the conservation of circulation is imposed
instead of that of momentum. They attribute this shortcoming of the
model to its over-simplicity. It is also not clear what physical mechanism
could account for the energy generated by the bore. In the SW framework,
energy generating bores are unphysical.

\citet{Klemp1994} solve the lock-exchange problem numerically by
using a characteristics-type method suggested by \citet{Rottman1983}.
This approach, however, differs significantly from the standard simple-wave
method \citep[Sec. 6.8]{Whitham1974} pursued in this study. \citet{Hogg2006}
uses a hodograph transform to solve the problem analytically in the
reduced gravity approximation. Effectively, this is a single-layer
solution, which is limited to large density differences. A direct
numerical solution of the problem has been attempted by \citet[(Sec. 2.4)][]{Ungarish2009}
using non-conservative two-layer SW equations. The discontinuous solutions
obtained in such a way are inherently spurious because they do not
satisfy relevant conservation laws. A lock-exchange problem with entrainment
has been modeled numerically by \citet{Milewski2015} using the two-layer
SW conservation laws for the circulation and energy and an advanced
finite-volume scheme. They also use a characteristics-type method
similar to that of \citet{Rottman1983} to obtain an analytical solution
to the problem with the conservation of either the mass or energy
besides that of circulation. Partial lock-exchange problem for Boussinesq
fluids has been solved numerically by \citet{Esler2011} using a weakly
non-hydrostatic SW approximation in which the wave dispersion prevents
the formation of sharp fronts. Recently, this problem was revisited
by \citet{Khodkar2017} using the so-called differential vorticity
model. Although the authors presume their model to be non-hydrostatic,
it produces the same characteristic equation as the hydrostatic two-layer
SW approximation. \citep{Rottman1983,Klemp1994} It implies that this
model is mathematically equivalent to the SW circulation conservation
law which, in turn, corresponds the singular limit $\alpha\rightarrow\infty$
of the generalized SW momentum equation considered in the present
study.

The paper is organized as follows. In the next section, we formulate
the problem and introduce a mathematical model based on the generalized
SW momentum equation. In section \ref{sec:analytic}, we use the method
of characteristics for simple waves to solve the problem analytically.
An alternative version of the lock exchange problem with a modified
initial state and the corresponding analytical solution are presented
in section \ref{sec:modprob}. In section \ref{sec:num}, the original
lock exchange problem is solved numerically using a composite Lax-Wendroff/Lax-Friedrichs
scheme to integrate locally conservative two-layer SW equations. The
paper is concluded with a summary and discussion of the main results
in section \ref{sec:Sum}.

\section{\label{sec:problem}Formulation of problem}

\begin{figure}
\begin{centering}
\includegraphics[width=0.6\columnwidth]{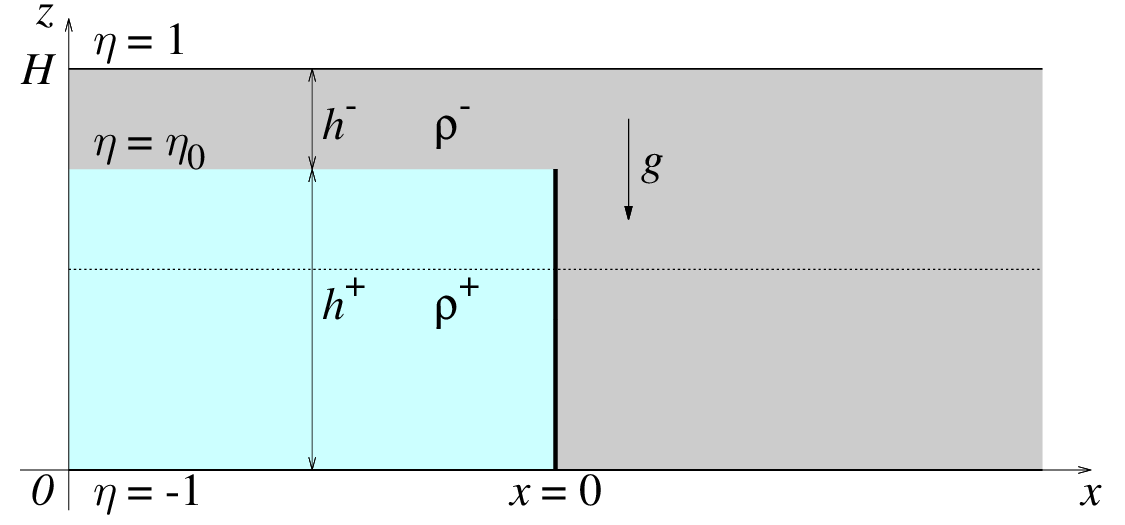}
\par\end{centering}
\caption{\label{fig:sketch}Sketch of the initial state of the lock-exchange
problem.}
\end{figure}

Consider a horizontal channel of constant height $H$ bounded by two
parallel solid walls and filled with two inviscid immiscible fluids
which are subject to a downward gravity force with the free fall acceleration
$g.$ Initially, a layer of heavier fluid of density $\rho^{+}$ and
uniform depth $h^{+}$ is overlaid by a lighter fluid of density $\rho^{-}$
and separated by a vertical lock gate from the same lighter fluid
on the right as shown in Fig. \ref{fig:sketch}. The lock is instantaneously
released and the heavier fluid starts to penetrate along the bottom
into the lighter fluid so forcing the latter in the opposite direction
at the top.

In the first-order (hydrostatic) SW approximation, which is based
on the assumption that the characteristic horizontal length scale
$L$ is much larger than the height $H:$ $H/L=\varepsilon\ll1,$
the mass conservation results in the vertical velocity component $w$
which is much smaller than the characteristic horizontal velocity
$u:$ $w/u\sim\varepsilon$. Such an effectively horizontal flow has
a dynamically negligible effect on the vertical pressure distribution
which is thus purely hydrostatic: 
\begin{equation}
p^{\pm}(x,z,t)=\mathit{\Pi}(x,t)-\rho^{\pm}g(z-h(x,t)).\label{eq:ppm}
\end{equation}
 The plus and minus indices refer to the bottom and top layers, respectively,
and $\mathit{\Pi}(x,t)=\left.p^{\pm}(x,z,t)\right|_{z=h}$ is the
pressure distribution along the interface. Substituting the pressure
distribution (\ref{eq:ppm}) into the Euler equation for the horizontal
velocity component $u^{\pm}$ in each layer yields the first SW equation,
while the second equation follows from the conservation of mass in
each layer: \citep{Pedlosky1979}
\begin{align}
\rho(u_{t}+uu_{x}\pm gh_{x}) & =-\mathit{\Pi}_{x},\label{eq:upm}\\
h_{t}+(uh)_{x} & =0.\label{eq:hpm}
\end{align}
The subscripts $t$ and $x$ stand for the corresponding partial derivatives
and the plus and minus indices at $\rho$, $u,$ and $h$ have been
dropped for the sake of brevity. Note that in the SW approximation,
which is a long-wave approximation, the relevant flow features with
the characteristic length scale comparable to the layer depth appear
as hydraulic jumps. 

The system of four SW equations (\ref{eq:upm},\ref{eq:hpm}) contains
five unknowns, $u^{\pm}$, $h^{\pm}$ and $\mathit{\Pi},$ and is
completed by adding the fixed height constraint $\{h\}\equiv h^{+}+h^{-}=H,$
which can be used to eliminate the top layer depth $h^{-}=H-h^{+}.$
Henceforth, the curly brackets denote the sum of the enclosed quantities.

Two more unknowns can be eliminated as follows. First, adding up the
mass conservation equations for both layers and using $\{h\}_{t}\equiv0,$
we obtain $\{uh\}=\Phi(t),$ which is the total flow rate. The channel
is assumed to be laterally closed. It corresponds to $\Phi\equiv0,$
and thus $u^{-}h^{-}=-u^{+}h^{+}.$ Second, the pressure gradient
$\mathit{\Pi}_{x}$ can be eliminated by subtracting Eq. (\ref{eq:upm})
for the top layer from that for the bottom layer. This leaves only
two unknowns, $U\equiv u^{+}h^{+}$ and $h=h^{+},$ and two equations,
which can be written in locally conservative form as follows 
\begin{align}
\left(\left\{ \rho/h\right\} U\right)_{t}+\left(\tfrac{1}{2}\left[\rho/h^{2}\right]U^{2}+g[\rho]h\right)_{x} & =0,\label{eq:u}\\
h_{t}+U_{x} & =0.\label{eq:h}
\end{align}
The square brackets above denote the difference of the enclosed quantities
between the bottom and top layers: $\left[f\right]\equiv f^{+}-f^{-}.$

In the following, the density difference is assumed to be small. According
to the Boussinesq approximation, this difference is important only
for the gravity of fluids, which drives the flow, but not for the
inertia of fluids. The problem is simplified further by using the
total height $H$ and the characteristic gravity wave speed $C=\sqrt{2Hg[\rho]/\{\rho\}}$
as the length and velocity scales, and $H/C$ as the time scale.

Then Eqs. (\ref{eq:u},\ref{eq:h}) take a remarkably symmetric form:
\citep{Milewski2015}
\begin{align}
\vartheta_{t}+\tfrac{1}{2}(\eta(1-\vartheta^{2}))_{x} & =0,\label{eq:vrt}\\
\eta_{t}+\tfrac{1}{2}(\vartheta(1-\eta^{2}))_{x} & =0,\label{eq:vlm}
\end{align}
where $\eta=[h]$ and $\vartheta=[u]$ are the dimensionless depth
and velocity differentials between the top and bottom layers. Subsequently,
the former is referred to as the interface height and the latter as
the shear velocity. The momentum and energy equations 
\begin{align}
(\eta\vartheta)_{t}+\tfrac{1}{4}(\eta^{2}+\vartheta^{2}-3\eta^{2}\vartheta^{2})_{x} & =0,\label{eq:mnt}\\
(\eta^{2}+\vartheta^{2}-\eta^{2}\vartheta^{2})_{t}+(\eta\vartheta(1-\eta^{2})(1-\vartheta^{2}))_{x} & =0,\label{eq:nrg}
\end{align}
can be obtained by multiplying Eq. (\ref{eq:vrt}) with $\eta$ and
$\eta^{2}\vartheta,$ respectively, and then using Eq. (\ref{eq:vlm})
to convert the resulting equations into locally conservative form.
Note that the conserved quantity $\eta\vartheta$ in Eq. (\ref{eq:mnt})
represents a pseudo-momentum (impulse). \citep{Priede2020} An infinite
sequence of further conservation laws can be constructed in a similar
way. \citep{Milewski2015} The generalized momentum equation can be
obtained by multiplying Eq. (\ref{eq:vrt}) with an arbitrary constant
$\alpha$ and adding to Eq. (\ref{eq:mnt}):

\begin{equation}
((\eta+\alpha)\vartheta)_{t}+\tfrac{1}{4}(\eta^{2}+\vartheta^{2}-3\eta^{2}\vartheta^{2}+2\alpha\eta(1-\vartheta^{2}))_{x}=0.\label{eq:gen}
\end{equation}
For more detailed derivation of Eqs. (\ref{eq:mnt})\textendash (\ref{eq:gen}),
we refer to \citet{Priede2020}. The constant $\alpha$ in Eq. (\ref{eq:gen}),
which defines the relative contribution of each layer to the pressure
gradient along the interface, is supposed to depend only on the ratio
of densities. For nearly equal densities, $\alpha\approx0$ is expected,
which corresponds to both layers affecting the pressure at the interface
with equal weight coefficients. In the following, we will obtain an
analytical solution for general $\alpha$ and then focus on three
particular cases: $\alpha=0,\infty,-1.$ The first two correspond
to the momentum and circulation conservation laws (\ref{eq:mnt})
and (\ref{eq:vrt}), whereas the third one reproduces the classical
front condition for gravity currents obtained by \citet{Benjamin1968}
as well as its generalization to internal bores by \citet{Klemp1997}.
The alternative front condition for internal bores proposed by \citet{Wood1984}
is reproduced by $\alpha=1$ but not considered here because it is
not applicable to gravity currents which are the key element of the
lock-exchange flow.

Equations (\ref{eq:vrt}) and (\ref{eq:vlm}) can be written in canonical
form as follows 
\begin{equation}
R_{t}^{\pm}+\lambda^{\pm}R_{x}^{\pm}=0\label{eq:canon}
\end{equation}
 using the characteristic velocities 
\begin{equation}
\lambda^{\pm}=\frac{3}{4}R^{\pm}+\frac{1}{4}R^{\mp}\label{eq:Cpm}
\end{equation}
and the Riemann invariants \citep{Long1956,Cavanie1969,Ovsyannikov1979,Sandstrom1993,Baines1995,Esler2011}
\begin{equation}
R^{\pm}=-\eta\vartheta\pm\sqrt{(1-\eta^{2})(1-\vartheta^{2})},\label{eq:Rpm}
\end{equation}
which are the constants of integration of the characteristic form
of Eqs. (\ref{eq:vrt}) and (\ref{eq:vlm}): 
\begin{equation}
\frac{\mathrm{d}\vartheta}{\mathrm{d}\eta}=\mp\sqrt{\frac{1-\vartheta^{2}}{1-\eta^{2}}}.\label{eq:char}
\end{equation}
Note that along two trajectories $x^{\pm}(t)$ defined in the $(x,t)$
plane by $\frac{dx^{\pm}}{dt}=\lambda^{\pm},$ Eq. (\ref{eq:canon})
with the respective index reduces to $\frac{dR^{\pm}}{dt}=0.$ It
means that the respective Riemann invariant is constant along that
trajectory which is referred to as the characteristic curve. \citep{Whitham1974}
This is the basic idea behind the method of characteristics which
we use to solve the problem analytically.

Since the interface is confined between the top and bottom boundaries,
which corresponds to $\eta^{2}\le1,$ the characteristic velocities
(\ref{eq:Cpm}) are real and, thus, the equations are of hyperbolic
type if $\vartheta^{2}\le1.$ The latter constraint on the shear velocity
is required for the stability of interface which would be otherwise
disrupted by a long-wave Kelvin-Helmholtz instability. \citep{Milewski2004}
It has to be noted that this instability is different to the usual
short-wave Kelvin-Helmholtz which is absent in the hydrostatic SW
approximation. \citep{Esler2011}

Integrating Eqs. (\ref{eq:vlm}) and (\ref{eq:gen}) across a discontinuity
at the point $x=\xi(t)$, where $\eta$ and $\vartheta$ have the
jumps $\left\llbracket \eta\right\rrbracket \equiv\eta_{+}-\eta_{-}$
and $\left\llbracket \vartheta\right\rrbracket \equiv\vartheta_{+}-\vartheta_{-}$
with the plus and minus subscripts denoting the corresponding quantities
in the front and back of the jump, and the double-square brackets
stand for the differentials of the enclosed quantities across the
jump, the jump propagation velocity can be expressed respectively
as follows \citep{Priede2020}
\begin{eqnarray}
\dot{\xi} & = & \frac{1}{2}\frac{\left\llbracket \vartheta(1-\eta^{2})\right\rrbracket }{\left\llbracket \eta\right\rrbracket },\label{eq:jmp1}\\
\dot{\xi} & = & \frac{1}{4}\frac{\left\llbracket \vartheta^{2}(1-2\eta^{2})+\eta(\eta+2\alpha)(1-\vartheta^{2})\right\rrbracket }{\left\llbracket (\eta+\alpha)\vartheta\right\rrbracket }.\label{eq:jmp2}
\end{eqnarray}
As for single layer, the jump conditions above consist of two equations
and contain five unknowns, $\eta_{\pm},$ $\vartheta_{\pm}$ and $\dot{\xi}.$
It means that two unknown parameters can be determined when the other
three are known.

As the jump conditions (\ref{eq:jmp1},\ref{eq:jmp2}) are non-linear,
multiple solutions are in principle possible. Some of these solutions
may be unphysical. For a solution to be feasible, it has to satisfy
the hyperbolicity constraint $\vartheta_{\pm}^{2}\le1$ as well as
the energy constraint which follows from the integration of Eq. (\ref{eq:nrg})
across the jump and defines the associated energy production rate:
\citep{Priede2020} 
\begin{equation}
\left\llbracket \eta\vartheta(1-\eta^{2})(1-\vartheta^{2})\right\rrbracket -\dot{\xi}\left\llbracket \eta^{2}+\vartheta^{2}-\eta^{2}\vartheta^{2}\right\rrbracket =\dot{\varepsilon}\le0.\label{eq:dsp}
\end{equation}
This quantity cannot be positive as the energy can only be dissipated
but not generated in the jump.

\section{\label{sec:analytic}Analytical simple-wave solution}

In this section, the lock-exchange problem will be solved analytically
using the simple-wave method \citep{Whitham1974} along with the analytic
expressions for the Riemann invariants (\ref{eq:Rpm}) and the associated
characteristic velocities (\ref{eq:Cpm}). To facilitate the solution
it is useful to introduce the following substitutions: 
\begin{eqnarray}
\eta & = & \sin\theta,\label{eq:theta}\\
\vartheta & = & \sin\phi,\label{eq:phi}
\end{eqnarray}
where $\theta$ and $\phi$ are the associated angular variables.
In these variables, the Riemann invariants and the characteristic
velocities can be written more concisely as follows 
\begin{align}
r^{\pm} & =\pm\arccos R^{\pm}=\phi\pm\theta,\label{eq:rpm}\\
\lambda^{\pm} & =\pm\frac{3}{4}\cos r^{\pm}\mp\frac{1}{4}\cos r^{\mp}.\label{eq:lpm}
\end{align}

The simple-wave method is applicable to the disturbances which propagate
into an initially homogeneous state. In this problem, there are two
such states: one on the right from the lock $(x>0)$ with the interface
located at the bottom $(\eta=-1)$ and the second on the left $(x<0)$
with the interface located at $\eta=\eta_{0}.$ Subsequently, these
regions will be referred to as the downstream and upstream states
according to the direction in which the heavier fluid flows.

\begin{figure}
\centering{}\includegraphics[width=0.5\columnwidth]{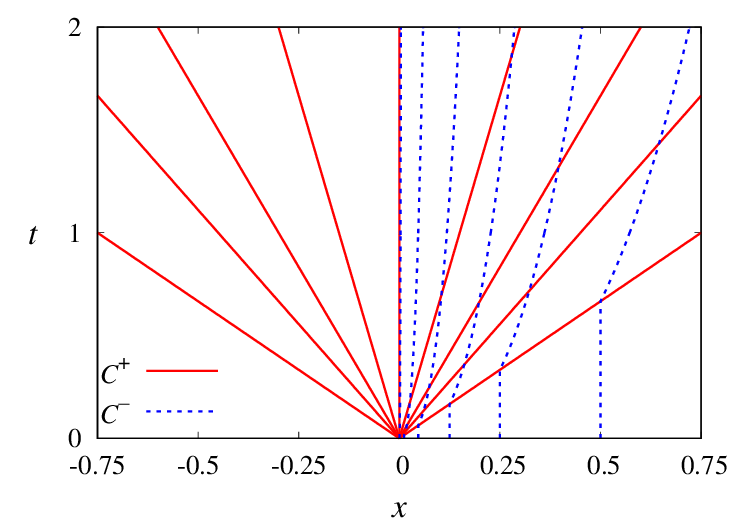}\put(-10,30){(a)}\includegraphics[width=0.5\columnwidth]{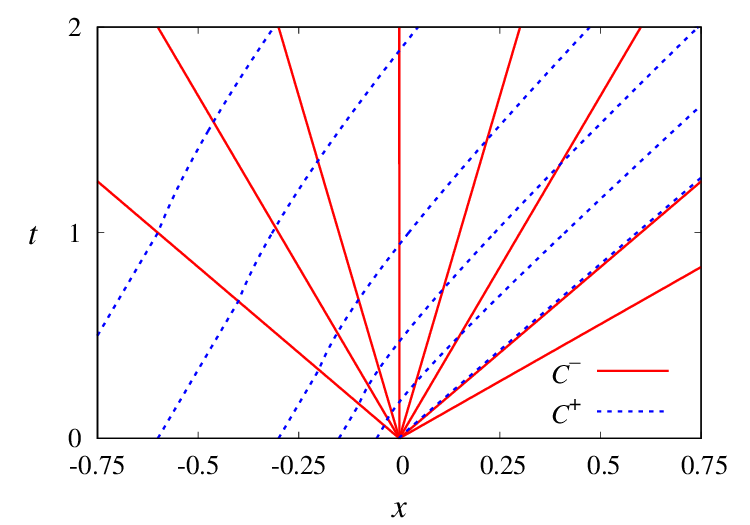}\put(-10,30){(b)}\caption{\label{fig:char}The $C^{+}$ and $C^{-}$ characteristics in the
downstream $(x>0)$ (a) and upstream $(x<0)$ (b) regions for the
partial lock exchange with $\eta_{0}=\cos\theta_{0}.$}
\end{figure}

Let us start with the downstream region, which is completely filled
with the lighter fluid, and consider the disturbances propagating
from the lock along the $C^{+}$ characteristics into this uniform
state where $(\eta,\vartheta)=(-1,0)$ and, correspondingly, $(\theta,\phi)=(-\frac{\pi}{2},0).$
Then the respective Riemann invariant along the $C^{-}$ characteristics,
which originate from this state, is $r^{-}=\phi-\theta=\frac{\pi}{2}.$
Applying this result to the $C^{+}$ characteristics, which propagate
from the lock into this state, this Riemann invariant can be written
as follows 
\[
r^{+}=\phi+\theta=\frac{\pi}{2}+2\theta,
\]
 and hence 
\begin{equation}
\lambda^{+}(\theta)=-\frac{3}{4}\sin2\theta=\frac{\mathrm{d}x}{\mathrm{d}t}.\label{eq:lmb+}
\end{equation}
 Since not only $r^{+}$ but also $r^{-}$ are invariant along $C^{+},$
owing to Eq. (\ref{eq:lpm}) the same holds also for $\lambda^{+}.$
Then Eq. (\ref{eq:lmb+}) can be integrated to obtain $\lambda^{+}=\frac{x}{t},$
which defines $C^{+}$ characteristics as the straight lines emanating
from the origin of the $(x,t)$ plane. This straightforwardly leads
to the solution, which can be written in parametric form as follows
\begin{align}
\frac{x}{t} & =-\frac{3}{4}\sin2\theta,\label{eq:xi-R}\\
\eta^{+} & =\sin\theta,\label{eq:eta-R}\\
\vartheta^{+} & =\cos\theta.\label{eq:v-R}
\end{align}
The $C^{-}$ slope, which is given by the corresponding characteristic
velocity and can be expressed as 
\begin{equation}
\lambda^{-}=-\frac{1}{4}\sin2\theta=\frac{1}{3}\lambda^{+},\label{eq:lmb-M}
\end{equation}
varies along the $C^{-}$ characteristics as they cross the $C^{+}$
characteristics. At $\theta=0,$ the slopes of both families of characteristics
become equal to each other: $\lambda^{+}=\lambda^{-}=0.$ It means
that the region where both families of characteristics intersect,
and thus solutions (\ref{eq:eta-R}) and (\ref{eq:v-R}) are applicable,
is limited to $-\frac{\pi}{2}\le\theta\le0.$ Equation (\ref{eq:lmb-M})
written in terms of $x$ and $t$ takes the form 
\[
\frac{\mathrm{d}x}{\mathrm{d}t}=\frac{1}{3}\frac{x}{t}
\]
and defines the $C^{-}$ characteristics above the rightmost $C^{+}$
characteristic, \emph{i.e.,} for $x\ge\frac{3}{4}t.$ The general
solution of this equation is $x(t)=ct^{1/3},$ where the unknown constant
$c$ can be determined by matching this solution with $x(t)=\text{const}$
for $0\le x\le\frac{3}{4}t,$ which corresponds to $\lambda^{\pm}=0$
for the undisturbed downstream state. Both families of characteristics
for the downstream region are shown in Fig. \ref{fig:char}(a).

\begin{figure}
\begin{centering}
\includegraphics[width=0.5\columnwidth]{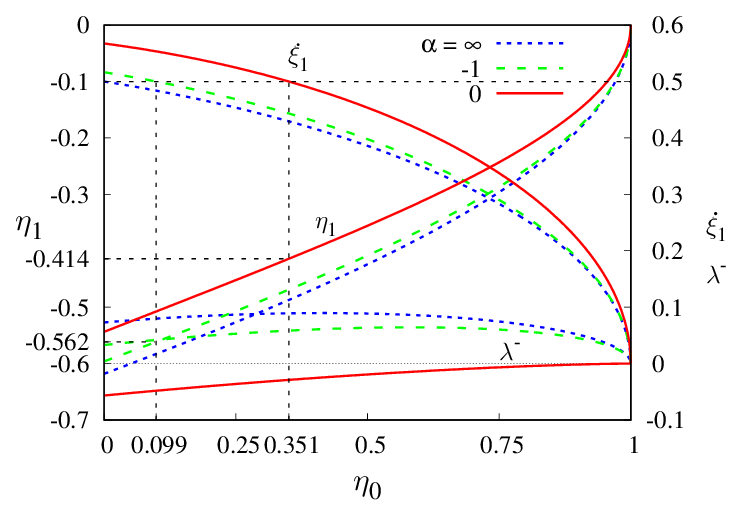}\put(-10,25){(a)}
\par\end{centering}
\centering{}\includegraphics[width=0.5\columnwidth]{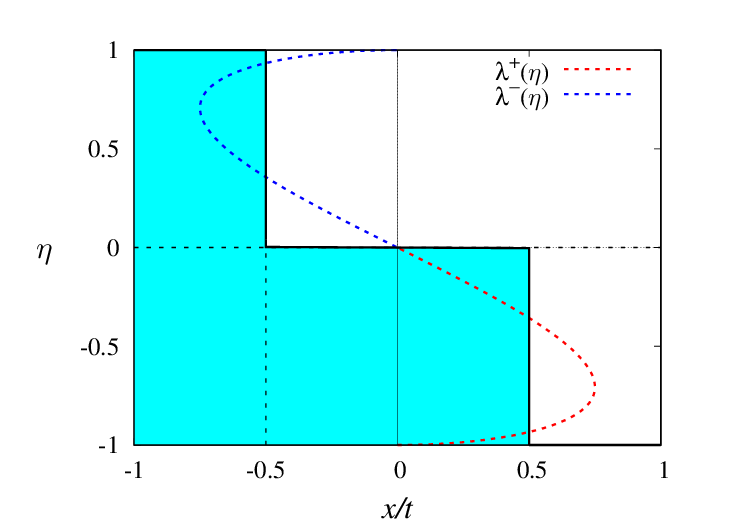}\put(-15,25){(b)}\includegraphics[width=0.5\columnwidth]{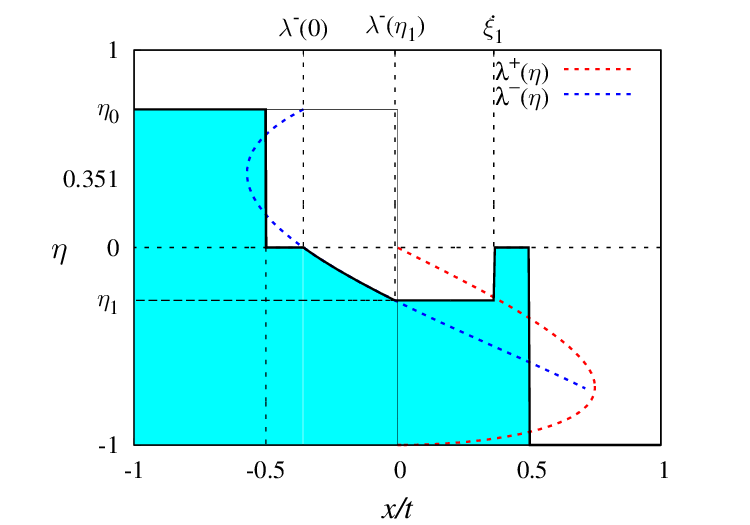}\put(-15,25){(c)}
\caption{\label{fig:hjmp}The interface height $\eta_{1},$ the velocity of
propagation $\dot{\xi}_{1}$ of the jump trailing the head-block and
the characteristic velocity $\lambda^{-}$ at $\eta=\eta_{1}$ versus
the upstream interface height $\eta_{0}$ (a); the interface height
versus the similarity variable $x/t$ for the full lock exchange:
$\eta_{0}=1$ (b) and partial lock exchange: $\eta_{c}<\eta_{0}<1,$
where $\eta_{c}=0.099$ is defined by Eq. (\ref{eq:etac}).}
\end{figure}

For the full lock exchange $(\eta_{0}=1),$ the problem is centrally
symmetric, hence the solution (\ref{eq:xi-R})\textendash (\ref{eq:v-R})
for $x>0$ holds also for $x<0.$ As seen in Fig. \ref{fig:hjmp}(b),
the respective interface height $\eta(x/t)$ is double valued. This
implies that the actual solution has to contain a jump.

Let us first consider $x>0$ and find the jump connecting solutions
(\ref{eq:eta-R},\ref{eq:v-R}) with the downstream state where $\eta_{+}=-1$
and $\vartheta_{+}=0.$ Substituting these downstream parameters into
the jump conditions (\ref{eq:jmp1},\ref{eq:jmp2}), after a few rearrangements,
we obtain 
\begin{eqnarray}
\vartheta^{2} & = & 1-\frac{2(1-\alpha)\eta}{\eta^{2}+2(\eta+\alpha)-1},\label{eq:v2-R}\\
\dot{\xi} & = & \frac{1}{2}(1-\eta)\vartheta,\label{eq:xid}
\end{eqnarray}
while Eqs. (\ref{eq:eta-R}, \ref{eq:v-R}) yield 
\begin{equation}
\vartheta^{2}=1-\eta^{2}.\label{eq:v2}
\end{equation}
 There are only two possible interface heights, $\eta=-1$ and $\eta=0,$
which satisfy Eqs. (\ref{eq:v2-R})\textendash (\ref{eq:v2}). The
former corresponds to the continuous double-valued solution whereas
the latter corresponds a jump spanning the lower half of the channel.
The possible shear velocities and the respective propagation speeds
for this jump, which are defined respectively by Eqs. (\ref{eq:v2-R},\ref{eq:xid}),
are $\vartheta=\pm1$ and $\dot{\xi}=\pm\frac{1}{2}.$ The downstream
direction of propagation corresponds to the positive solution. The
velocity of propagation being independent of $\alpha$ implies that
this jump conserves not only the mass and impulse but also the circulation
and energy. \citep{Priede2020}

For the full lock exchange, there is a centrally symmetric upstream
jump at $x<0,$ which spans the upper half of the channel from $\text{\ensuremath{\eta}}=0$
to $\eta=1$ and moves leftward at the velocity $\dot{\xi}=-\frac{1}{2}$
(see Fig. \ref{fig:hjmp}b). This exact fully-conservative solution
coincides with that assumed by \citet{Yih1955} but differs from the
numerical solution obtained by \citet{Klemp1994} which will be considered
later in connection with the possible instability of deep gravity
currents. \citep{Priede2020} 

Let us now consider a partial lock exchange with the upstream interface
height $\eta_{0}=\sin(\frac{\pi}{2}-\theta_{0}),$ where $0\le\theta_{0}\le\pi.$
This initial state with $\eta=\eta_{0}$ and $\vartheta=0$ corresponds
to the angular variables $\theta=\frac{\pi}{2}-\theta_{0}$ and $\phi=0.$
Then the positive Riemann invariant takes the form
\[
r^{+}=\theta+\phi=\frac{\pi}{2}-\theta_{0},
\]
from where we have $\phi=\frac{\pi}{2}-\theta_{0}-\theta.$ Using
this relation, the negative Riemann invariant associated with the
$C^{-}$ characteristics, which extend upstream from the lock, can
be written as 
\[
r^{-}=\phi-\theta=\frac{\pi}{2}-\theta_{0}-2\theta.
\]
Since the respective characteristic velocity (\ref{eq:lpm}) 
\begin{equation}
\lambda^{-}=-\frac{3}{4}\sin(2\theta+\theta_{0})+\frac{1}{4}\sin\theta_{0}\label{eq:lmb-}
\end{equation}
 is constant along $C^{-},$ these characteristics are straight lines
with the slope $\lambda^{-}=\frac{x}{t}.$ Consequently, the solution
can be written parametrically as follows: 
\begin{align}
\frac{x}{t} & =-\frac{3}{4}\sin(2\theta+\theta_{0})+\frac{1}{4}\sin\theta_{0},\label{eq:xi-L}\\
\eta^{-} & =\sin\theta,\label{eq:eta-L}\\
\vartheta^{-} & =\cos(\theta+\theta_{0}).\label{eq:v-L}
\end{align}
To determine the range of applicability of this solution, we need
to consider also the $C^{+}$ characteristics. The slope of these
characteristics varying across the $C^{-}$ as follows
\begin{equation}
\lambda^{+}=\frac{3}{4}\sin\theta_{0}-\frac{1}{4}\sin(2\theta+\theta_{0})=\frac{1}{3}\lambda^{-}+\frac{2}{3}\sin\theta_{0}.\label{eq:lmb-P}
\end{equation}
becomes equal at $\theta=-\theta_{0}$ to that of $C^{-}:$ $\lambda^{+}=\lambda^{-}=\sin\theta_{0}\ge0.$
At this point, both families of characteristics become parallel to
each other. It means that the applicability of solution (\ref{eq:xi-L})\textendash (\ref{eq:v-L})
is limited to $-\theta_{0}\le\theta\le\frac{\pi}{2}-\theta_{0}.$
Equation (\ref{eq:lmb-P}) defining the $C^{+}$ characteristics can
be written in terms of $x$ and $t$ as follows 
\[
\frac{\mathrm{d}x}{\mathrm{d}t}=\frac{1}{3}\frac{x}{t}+\frac{2}{3}\sin\theta_{0}.
\]
 The general solution of this equation is $x(t)=ct^{1/3}+t\sin\theta_{0},$
where $c$ is an unknown constant. The latter can be determined by
matching this solution with that for the undisturbed upstream state
$x(t)=x_{0}+\frac{1}{2}t\sin\theta_{0},$ which holds below the leftmost
$C^{-}$characteristic defined by 
\[
\frac{x}{t}\le\min\lambda^{-}=\frac{1}{3}\sin\theta_{0}-\frac{4}{3}.
\]
Both families of characteristics for the upstream region are shown
in Fig. \ref{fig:char}(b).

Downstream from the lock $(x>0),$ where the initial state is the
same as for the full lock exchange, solution remains defined by Eqs.
(\ref{eq:xi-R})\textendash (\ref{eq:v-R}), which hold for $-\frac{\pi}{2}\le\theta<0.$
Note that for $\theta_{0}>0,$ solution (\ref{eq:xi-L})\textendash (\ref{eq:v-L})
extends downstream up to $\frac{x}{t}=\sin\theta_{0},$ which corresponds
to $\theta=-\theta_{0}$ in Eq. (\ref{eq:xi-L}). Thus, this solution
overlaps with that defined by Eqs. (\ref{eq:xi-R})\textendash (\ref{eq:v-R})
in the sector $0\le\frac{x}{t}\le\sin\theta_{0}.$ As usual, the double-valuedness
of this solution implies the presence of a jump.

Let us first consider the jump connecting the upstream state $\eta=\eta_{0}$
and $\vartheta=0$ with solution (\ref{eq:xi-L})\textendash (\ref{eq:v-L}).
Substituting solution (\ref{eq:eta-L},\ref{eq:v-L}) into the jump
conditions (\ref{eq:jmp1},\ref{eq:jmp2}), we obtain:
\begin{equation}
\vartheta^{2}=\frac{(\eta_{0}-\eta)^{2}(\eta_{0}+\eta+2\alpha)}{(1-\eta^{2})(\eta_{0}-\eta)+2(\eta+\alpha)(1-\eta_{0}\eta)}=\left(\eta_{0}\sqrt{1-\eta^{2}}-\eta\sqrt{1-\eta_{0}^{2}}\right)^{2}.\label{eq:v2-L}
\end{equation}
This equation has only two roots: $\eta=\eta_{0}$ and $\eta=0.$
As for the full lock exchange, the first root corresponds to the original
continuous solution, which however is double valued if $\eta_{0}>0.$
The second root describes a jump from $\eta=\eta_{0}$ to the channel
mid-height $\eta=0$. The shear velocity behind this jump and its
velocity of propagation, which are defined respectively by Eqs. (\ref{eq:v2-L})
and (\ref{eq:jmp1}), are $\vartheta=\eta_{0}$ and $\dot{\xi}=-\frac{1}{2}\frac{\vartheta}{\eta_{0}}=-\frac{1}{2}.$
The fact that $\dot{\xi}$ is independent of $\eta_{0}$ as well as
of $\alpha$ implies that this jump is fully conservative and, thus,
represents a long-wave approximation to the so-called solibore which
appears in the non-hydrostatic model. \citep{Priede2020} 

In the downstream direction, the uniform state behind this jump $(\eta,\vartheta)=(0,\eta_{0})$
connects with the solution (\ref{eq:xi-L})\textendash (\ref{eq:v-L})
at the point $\theta=0.$ According to Eq. (\ref{eq:xi-L}), this
point moves at the velocity $\frac{x}{t}=-\frac{1}{2}\sin\theta_{0}\le-\frac{1}{2},$
which does not exceed the velocity of the upstream jump. At larger
distances $x/t,$ the interface, which is defined parametrically by
Eqs. (\ref{eq:xi-L},\ref{eq:eta-L}), descents below the mid-height
$\eta=0.$ This solution can be connected with the solution downstream
from the lock which, as noted above, is the same as that for the full
lock exchange: $\eta_{+}=0$ and $\vartheta_{+}=1.$ With this front
state, the jump conditions (\ref{eq:jmp1},\ref{eq:jmp2}) yield the
following relationship between $\vartheta_{-}=\vartheta_{1}$ and
$\eta_{-}=\eta_{1}$ behind the jump: 
\[
\vartheta_{1}=\frac{(2\alpha+\eta_{1})(1-\eta_{1}^{2})}{2\alpha+\eta_{1}(1+\eta_{1}^{2})}.
\]
 Substituting solution (\ref{eq:eta-L},\ref{eq:v-L}) into this equation,
we obtain 
\begin{equation}
\vartheta_{1}=\cos(\theta_{1}+\theta_{0})=\frac{(2\alpha+\sin\theta_{1})(1-\sin^{2}\theta_{1})}{2\alpha+\sin\theta_{1}(1+\sin^{2}\theta_{1})},\label{eq:v1-2}
\end{equation}
which relates $\theta_{1}$ behind the jump with the corresponding
upstream quantity $\theta_{0}=\arccos\eta_{0}.$ This equation is
solvable analytically for $\theta_{0}$ in terms of $\theta_{1}$
or numerically the other way round. The analytical solution $\eta_{1}=\sin\theta_{1}$
for the interface height behind the jump and its velocity of propagation
\begin{equation}
\dot{\xi}_{1}=-\frac{\eta_{1}(3-\eta_{1}^{2})}{2(1+\eta_{1}^{2})}\label{eq:xid1}
\end{equation}
as well as the characteristic velocity $\lambda^{-}(\eta_{1})$ are
plotted in Fig. \ref{fig:hjmp}(a) against the upstream interface
height $\eta_{0}.$ As seen in Fig. \ref{fig:hjmp}(c), $\lambda^{-}(\eta_{1})$
defines the upstream limiting point of the flat depressed interface
region which forms behind the head block with the trailing edge advancing
at a supercritical velocity $\dot{\xi}_{1}>\lambda^{-}(\eta_{1}).$
Figure \ref{fig:hjmp}(a) shows that $\dot{\xi}_{1}$ increases with
decreasing $\eta_{0}$ and attains the velocity of the leading head
block edge $\dot{\xi}=0.5$ at 
\begin{equation}
\eta_{c}\approx0.351,\,0.099,\,0\quad\text{for}\quad\alpha=0,-1,\infty.\label{eq:etac}
\end{equation}
At this critical point, both edges merge and, thus, the head block
vanishes. It means that for $\eta_{0}\le\eta_{c},$ solution (\ref{eq:xi-L})\textendash (\ref{eq:v-L})
has to connect directly to the quiescent downstream state with $\eta_{+}=-1$
and $\vartheta_{+}=0.$ Then the interface height and the shear velocity
behind the jump are related by Eq. (\ref{eq:v2-R}). Combining this
relation with solution (\ref{eq:eta-L},\ref{eq:v-L}), we obtain

\begin{equation}
\vartheta_{1}=\cos(\theta_{1}+\theta_{0})=\left(\frac{\cos^{2}\theta_{1}-2\alpha(1+\sin\theta_{1})}{\cos^{2}\theta_{1}-2(\alpha+\sin\theta_{1})}\right)^{1/2}.\label{eq:v1-1}
\end{equation}
Similar to Eq. (\ref{eq:v1-2}), this equation is solvable analytically
for $\theta_{0}$ in terms of $\theta_{1}$ or numerically the other
way round. The speed of propagation $\dot{\xi}$ is obtained by substituting
$\eta_{1}=\sin\theta_{1}$ and $\vartheta_{1}$ found by solving this
equation for $\eta$ and $\vartheta$ in Eq. (\ref{eq:xid}).

\begin{figure}
\begin{centering}
\includegraphics[width=0.5\columnwidth]{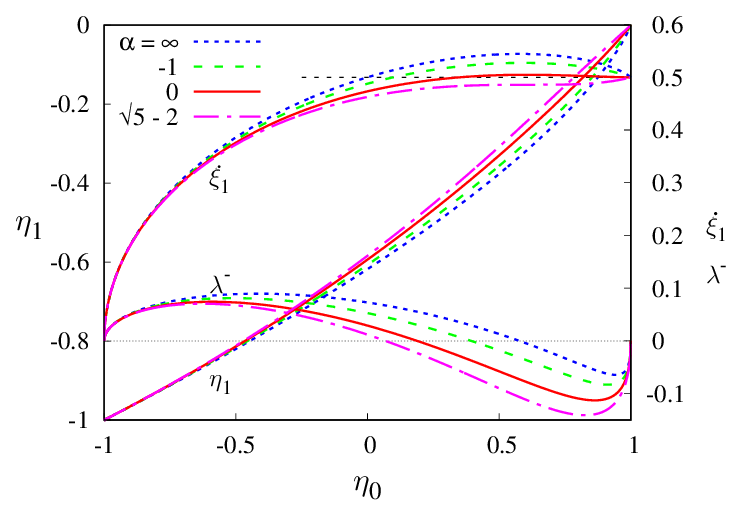}\put(-10,25){(a)}
\par\end{centering}
\centering{}\includegraphics[width=0.5\columnwidth]{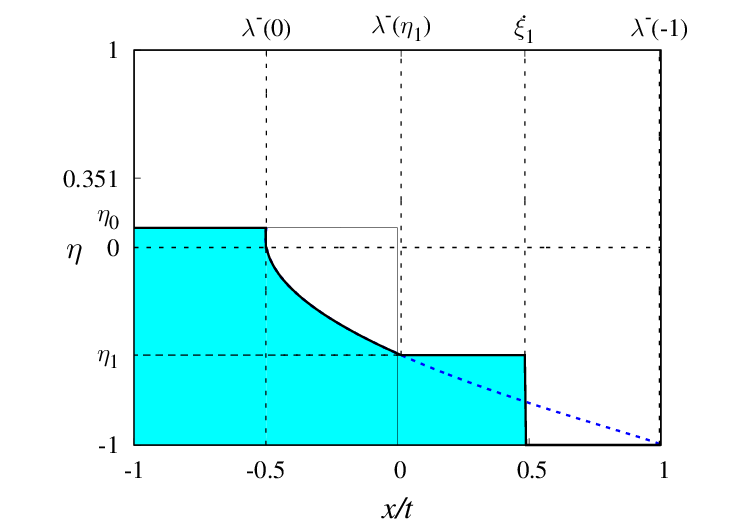}\put(-15,25){(b)}\includegraphics[width=0.5\columnwidth]{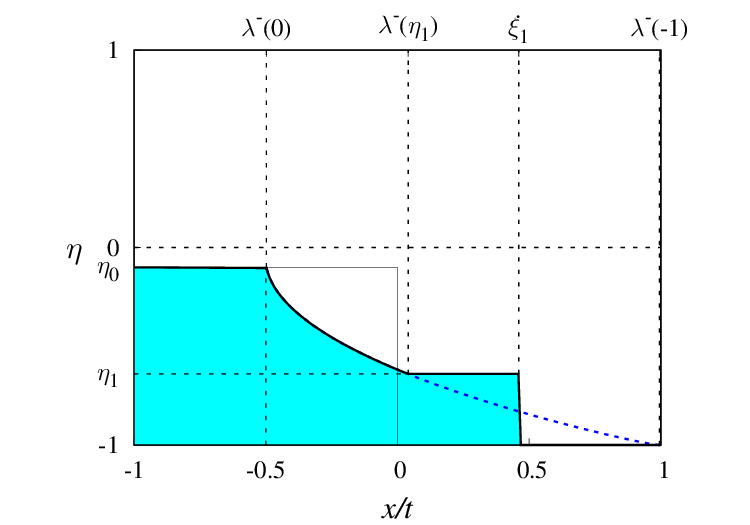}\put(-15,25){(c)}\caption{\label{fig:hjmp1}The interface height and the velocity of propagation
of the leading jump (a) and the overall interface shape for the partial
lock exchange with $0\le\eta_{0}\le0.351$ (b) and $\eta_{0}\le0$
(c).}
\end{figure}

The downstream jump parameters defined by Eq. (\ref{eq:v1-1}) are
plotted in Fig. \ref{fig:hjmp1}(a) for the whole range of upstream
heights which, in principle, admit this type of solution without the
head block. The overall interface height versus the similarity variable
$x/t$ is shown in Figs. \ref{fig:hjmp1}(b,c). As seen, there are
two different interface configurations possible in this case. The
first configuration, which is shown in Fig. \ref{fig:hjmp1}(b), corresponds
to $0\le\eta_{0}\le\eta_{c}$ and features an upstream jump from $\eta_{0}$
to the mid-height $\eta=0$ as in the case with the head block. The
second configuration corresponding to $\eta_{0}\le0$, which is shown
in Fig. \ref{fig:hjmp1}(c), has no upstream jump. In the latter case,
the upstream state connects directly to solution (\ref{eq:eta-L},\ref{eq:v-L})
at $\eta=\eta_{0},$ which is the other root of Eq. (\ref{eq:v2-L}).

It is important to note that for $\eta_{0}\le0,$ solution (\ref{eq:eta-L},\ref{eq:v-L})
can in principle connect directly to the downstream state $\eta=-1$
at $\theta=-\frac{\pi}{2}$ without a leading jump, as in the dam-break
solution for the single-layer case. The following considerations,
however, imply that, in the two-layer case, this alternative solution
is inherently unstable. Namely, as seen in Fig. \ref{fig:hjmp1}(b,c),
the sharp edge in the continuous solution propagates with a non-zero
characteristic velocity $\lambda^{-}$ which is defined by Eq. (\ref{eq:lmb-})
with $\theta=-\pi/2$ corresponding to $\eta=-1.$ Now consider a
perturbation resulting in a small non-zero edge height. As it may
be seen in Fig. \ref{fig:hjmp1}(a), the speed of propagation $\dot{\xi}$
for jump of finite height drops to zero in this limit. This effectively
halts the propagation of the edge causing its height to increase further
as long as the fluid behind it moves faster than the edge. As a result,
the edge acquires a finite height. This is confirmed by the subsequent
direct numerical solution. A similar mechanism can be behind the formation
of the elevated head block when $\eta_{0}>\eta_{c}$.

For a thin bottom layer with $h_{0}=\frac{1}{2}(1+\eta_{0})\rightarrow0,$
Eq. (\ref{eq:v1-1}) can be solved explicitly as follows 
\[
\theta_{1}+\pi/2\approx(\pi-\theta_{0})/(1+1/\sqrt{2}).
\]
Using the above result, we readily recover the solution for the single-layer
dam-break problem with the reduced gravity and von Kármán front condition:
$\dot{\xi}_{1}=\sqrt{2h_{1}}$ and $\lambda^{+}=(\sqrt{2}-1)\sqrt{h_{1}},$
where $h_{1}=2(\sqrt{2}-1)^{2}h_{0}\approx0.343h_{0}$ is the downstream
front height. \citep[Sec. 2.5]{Ungarish2009} As noted above, a finite
front height is caused by the instability of the sharp edge in the
two-layer dam-break solution.

To conclude this section let us note that the existence of analytical
expressions for the Riemann invariants used above is advantageous
but not crucial for solving the problem since the characteristic equation
(\ref{eq:char}) can also be integrated numerically as done, for example,
by \citet{Klemp1994} and \citet{Khodkar2017} using the long form
of this equation. \citep{Rottman1983} In this case, the uniform initial
states downstream and upstream of the lock, $\left.\vartheta\right|_{\eta=-1}=0$
and $\left.\vartheta\right|_{\eta=\eta_{0}}=0,$ provide boundary
conditions for Eq. (\ref{eq:char}). Since this equation represents
two first-order ordinary differential equations defined by the plus
and minus signs at the second term, only one boundary condition is
required for each equation. Because the sign in Eq. (\ref{eq:char})
determines the direction of flow, it depends on which side from the
lock gate the heavier fluid is contained. For the configuration considered
in this study with the heavier fluid contained on the left, the downstream
and upstream boundary conditions apply to Eq. (\ref{eq:char}) with
the plus sign and minus signs, respectively. 

\section{\label{sec:modprob}Lock exchange with a modified initial state}

As already noted, the solution found above differs from the previously
obtained numerical solution. \citep{Klemp1994} The difference is
due to the non-standard method of characteristics used by \citet{Klemp1994}
who employ Benjamin's front condition directly as a boundary condition
for the characteristic equation. In the conventional simple-wave approach
used in the previous section, it is the quiescent downstream state
which defines the relevant boundary condition. The front condition,
in turn, is used to eliminate multi-valued parts of the solution,
if any, by fitting in the jumps. These two approaches are not equivalent
and lead to different results because they describe different physical
problems. 

Using the front condition as a boundary condition is equivalent to
assuming the initial downstream state to be a gravity current rather
than a quiescent fluid as in the original lock-exchange problem. Such
an assumption may be justified if the loss of hyperbolicity or instability
breaks the dependence of the solution on the initial state. In this
case, one can expect a stable gravity current of a lower height to
form as the effective downstream state. Although the height of this
gravity current is not uniquely defined, the problem can still be
solved analytically in a similar way to the original lock-exchange
problem in the previous section. For the sake of completeness, in
this section, we present also an analytical solution of this alternative
formulation.

\begin{figure}
\centering{}\includegraphics[width=0.5\columnwidth]{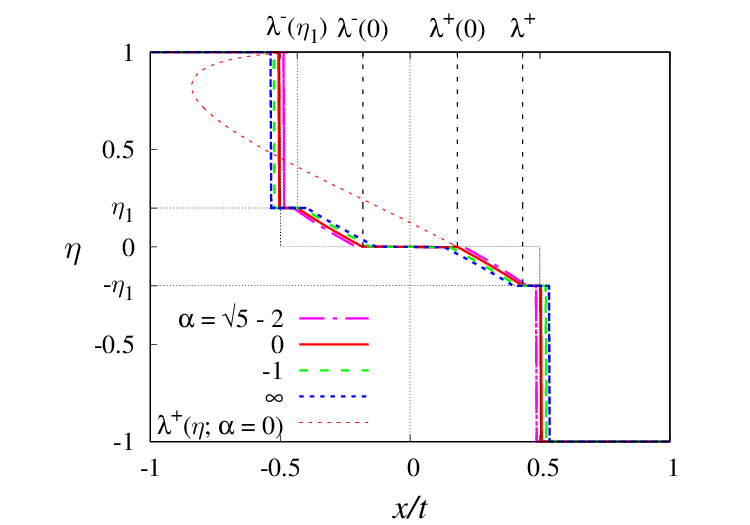}\put(-15,25){(a)}\includegraphics[width=0.5\columnwidth]{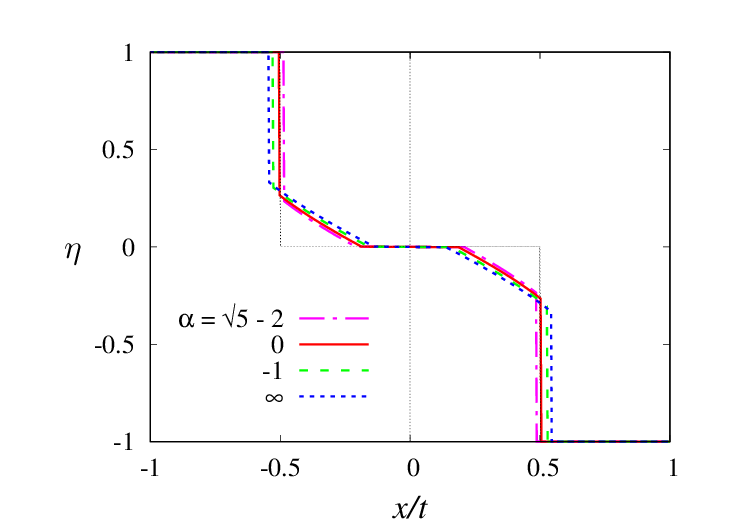}\put(-15,25){(b)}\caption{\label{fig:char-1}The interface height in the modified lock-exchange
problem for various $\alpha$ with the upstream and downstream gravity
current heights fixed to $\eta_{0}=\pm0.2$ (a) and with the lowest
possible gravity gravity current height for each $\alpha$ (b).}
\end{figure}

\begin{figure}
\centering{}\includegraphics[width=0.5\columnwidth]{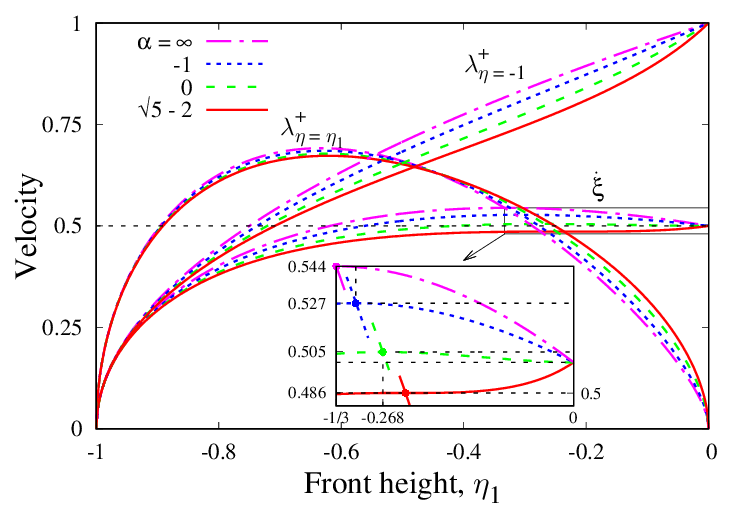}\caption{\label{fig:hcrt}The characteristic downstream wave speed $\lambda^{+}$
at the top $(\eta=\eta_{1})$ and bottom $(\eta=-1)$ of the gravity
current and the front velocity $\dot{\xi}$ versus the front height
for $\alpha=\infty,-1,0,\sqrt{5}-2.$}
\end{figure}

Following \citet{Klemp1994}, we now assume the downstream state to
be a gravity current with the interface height $\eta_{1}$ and the
shear velocity $\vartheta_{1}(\eta_{1})\ge0$ defined by Eq. (\ref{eq:v2-R}).
Then the negative Riemann invariant (\ref{eq:rpm}), which is constant
along the characteristics propagating upstream from this uniform state,
can be written as follows 
\begin{equation}
r^{-}=\phi-\theta=\arcsin\vartheta_{1}(\eta_{1})-\arcsin\eta_{1}.\label{eq:r-}
\end{equation}
Rearranging this expression as $\phi=r^{-}+\theta,$ the positive
Riemann invariant can be written as follows 
\[
r^{+}=\phi+\theta=r^{-}+2\theta.
\]
Since both $r^{-}$ and $r^{+}$ are constant along the $C^{+}$ characteristics,
which propagate into the downstream state, so is also the associated
characteristic velocity (\ref{eq:lpm}) 
\[
\lambda^{+}=\frac{3}{4}\cos r^{+}-\frac{1}{4}\cos r^{-}.
\]
 Then the solution can be written in the parametric form as follows
\begin{align}
\frac{x}{t} & =\frac{3}{4}\cos(r^{-}+2\theta)-\frac{1}{4}\cos r^{-},\label{eq:xi-Ra}\\
\eta^{+} & =\sin\theta,\label{eq:eta-Ra}\\
\vartheta^{+} & =\sin(r^{-}+\theta),\label{eq:v-Ra}
\end{align}
where $\eta\ge\eta_{1}$ and $r^{-}$ depends on $\eta_{1}$ as defined
by Eq. (\ref{eq:r-}). For the full lock exchange, the symmetry of
the problem implies: $\{\eta,\vartheta\}(-x/t)=\{-\eta,\vartheta\}(x/t).$
It means that the upstream and downstream solutions connect at $\eta=0$
without a jump, as shown in Fig. \ref{fig:char-1}(a) for $\eta_{1}=-0.2$
and various $\alpha.$ For such a solution to be possible, the gravity
current cannot propagate slower than the downstream characteristic
velocity for the respective height: $\dot{\xi}(\eta_{1})\ge\lambda^{+}(\eta_{1}),$
where the latter defines the speed at which the lower point of the
sloped part of the interface moves. This, in turn, implies that the
height of gravity current $\eta_{1}$ cannot be lower than the critical
value $\eta_{c}$ which is defined depending on $\alpha$ by $\dot{\xi}(\eta_{c})=\lambda^{+}(\eta_{c}).$
As seen in Fig. \ref{fig:hcrt}, $\eta_{c}$ coincides with the point
at which $\dot{\xi}(\eta_{1})$ attains maximum for a given $\alpha$
provided that $\alpha<\alpha_{c}=\sqrt{5}-2.$ \citep{Priede2020}
If $\alpha$ exceeds $\alpha_{c},$ the critical height $\eta_{c}$
switches from the maximum to the minimum of $\dot{\xi}.$ The latter
emerges at $\eta=-\alpha$ when $\alpha$ becomes greater than zero
and descends towards the maximum with the increase in $\alpha.$ At
$\alpha=\alpha_{c}$, both stationary points merge forming an inflection
point. As a result, the front velocity $\dot{\xi}$ becomes a monotonically
increasing function of $\eta_{1}$ at this particular value of $\alpha.$
If $\alpha$ exceeds $\alpha_{c},$ the inflection points splits into
two stationary points: a maximum located at $\eta_{c}=-\alpha$ and
minimum moving back towards the mid-plane $\eta=0$ which it reaches
at $\alpha=\frac{1}{2}.$

\begin{figure}
\centering{}\includegraphics[width=0.5\columnwidth]{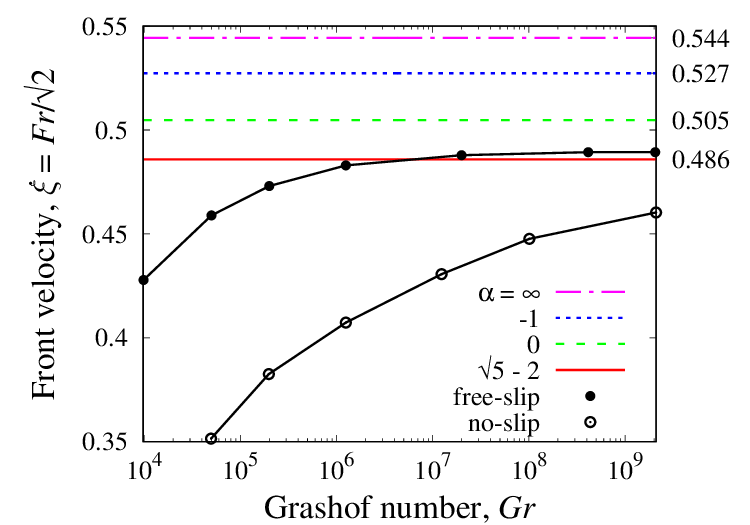} \caption{\label{fig:lckx}Comparison of critical gravity current speeds for
various $\alpha$ with the numerical results of \citet{Haertel2000}
for gravity currents generated by the lock exchange with free-slip
and no-slip boundary conditions. The conversion factor of $1/\sqrt{2}$
is due to the channel half-height used as the length scale in the
definition of Froude number $\mathit{Fr}$ by \citet{Haertel2000}.
The Grashof number defines the magnitude of the driving force which
produces characteristic flow velocity with the Reynolds number $\mathit{Re}\sim\sqrt{\mathit{Gr}}.$}
\end{figure}

The considerations above imply that if the initial height of gravity
current $\eta_{1}$ is lower than $\eta_{c},$ the fluid beneath the
slopped interface would run over the front until its height reaches
$\eta_{c}$ for a given $\alpha.$ The highest possible $\eta_{c}$
is attained at $\alpha=\alpha_{c}.$ It is because for $\alpha>\alpha_{c}$
the critical height $\eta_{c}$ switches to the local minimum of $\dot{\xi}$
but the latter is expected to be unstable. \citep{Benjamin1968,Baines2016}
Namely, a virtual perturbation decreasing the front height would increase
the front speed. Then the mass conservation would enhance the initial
perturbation and, thus, cause the gravity current collapse to a lower
depth. Therefore, $\eta_{c}=-\alpha_{c}$ is the highest possible
value of $\eta_{c}.$ The corresponding front velocity $\dot{\xi}=\alpha_{c}^{1/2}\approx0.486$
can be seen in Fig. \ref{fig:lckx} to agree very well with the highly
accurate numerical results of \citet{Haertel2000} for the gravity
currents generated by the lock exchange with free-slip boundary conditions.
The other values of $\alpha$ produce noticeable higher front velocities.
With real no-slip boundary conditions, a much higher Reynolds number
seems to be required to achieve this inviscid limit.

The solution obtained above can easily be extended to the partial-height
lock exchange where the upstream state is a layer of quiescent fluid
with the interface located at $\eta=\eta_{0}.$ In this case, the
initial upstream state is supposed to contain a bore with the interface
height $\eta_{1}\le\eta_{0}$ and the shear velocity $\vartheta_{1}(\eta_{0},\eta_{2})$
which is defined by the LHS of Eq. (\ref{eq:v2-L}) with $\eta_{1}$
standing instead of $\eta.$ By the same arguments as before, we find
that solution is described by Eqs. (\ref{eq:xi-L})\textendash (\ref{eq:v-L})
with $\theta_{0}=\frac{\pi}{2}-r^{+},$ where 
\begin{equation}
r^{+}=\phi+\theta=\arcsin\vartheta_{1}(\eta_{0},\eta_{1})+\arcsin\eta_{1}\label{eq:r+}
\end{equation}
 is the positive Riemann invariant corresponding to the upstream bore.

\begin{figure}
\centering{}\includegraphics[width=0.5\columnwidth]{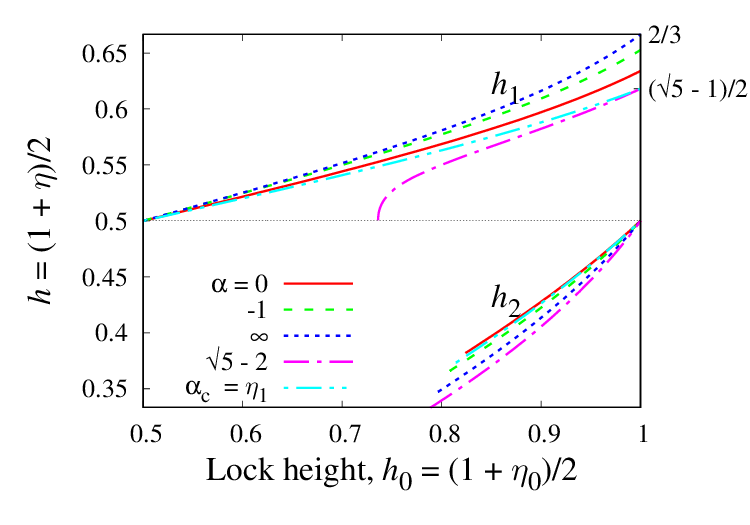}\put(-20,25){(a)}\includegraphics[width=0.5\columnwidth]{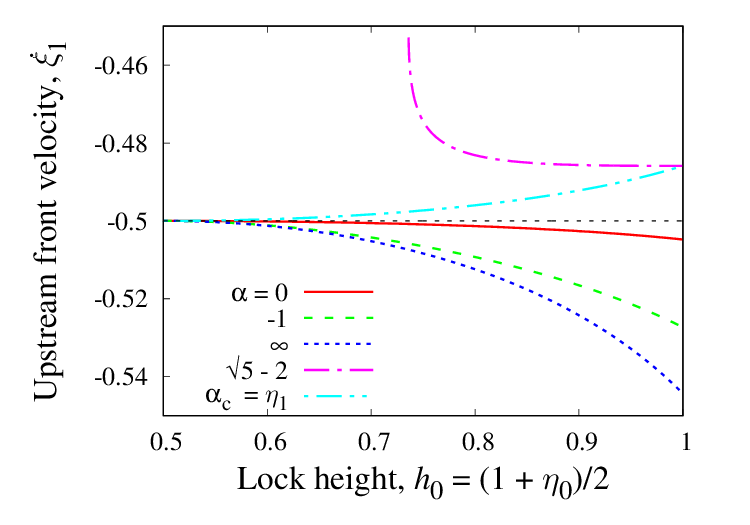}\put(-15,25){(b)}
\caption{\label{fig:lckx-mod-h01}The interface height $h_{1}$ that produces
the highest stable bore, the intermediate interface height $h_{2}$
(a) and the respective propagation velocity (b) against the lock height
$h_{0}$ for various $\alpha.$}
\end{figure}

The lowest possible bore height $h_{1}=(1+\eta_{1})/2,$ which is
determined by the corresponding characteristic velocity becoming equal
with the maximal front velocity for a given $\alpha,$ is plotted
in Fig. \ref{fig:lckx-mod-h01}(a) against the lock height $h_{0}$
for various $\alpha.$ For $\alpha>0,$ the energy constraint (\ref{eq:dsp}),
which can be written as follows 
\[
\dot{\varepsilon}=\frac{\eta_{1}\left(\alpha-\eta_{0}\right)\left(1-\eta_{1}^{2}\right)\left(\eta_{0}-\eta_{1}\right){}^{3}\sqrt{2\alpha-\eta_{0}-\eta_{1}}}{\left(3\eta_{0}\eta_{1}^{2}-\eta_{1}^{3}+2\alpha(1-\eta_{0}\eta_{1})-\eta_{1}-\eta_{0}\right){}^{3/2}},
\]
permits such upstream bores only for $\eta_{0}\ge2\alpha.$ This
constraint can be relaxed by assuming $\alpha$ to vary depending
on $\eta_{0}$ so as to minimize the maximum of propagation velocity
for a given $\eta_{0}.$ As before, this happens when the maximum
of $\dot{\xi}$ merges with the minimum forming a stationary inflection
point at $\eta_{1}=\alpha\ge0.$ The solution of $\partial_{\eta_{1}}^{2}\left.\dot{\xi}(\eta_{0},\eta_{1};\alpha)\right|_{\eta_{1}=\alpha}=0$
defining this critical point, which can be found analytically but
not presented here due to its complexity, is plotted in Fig. \ref{fig:lckx-mod-h01}(a).

\begin{figure}
\centering{}\includegraphics[width=0.5\columnwidth]{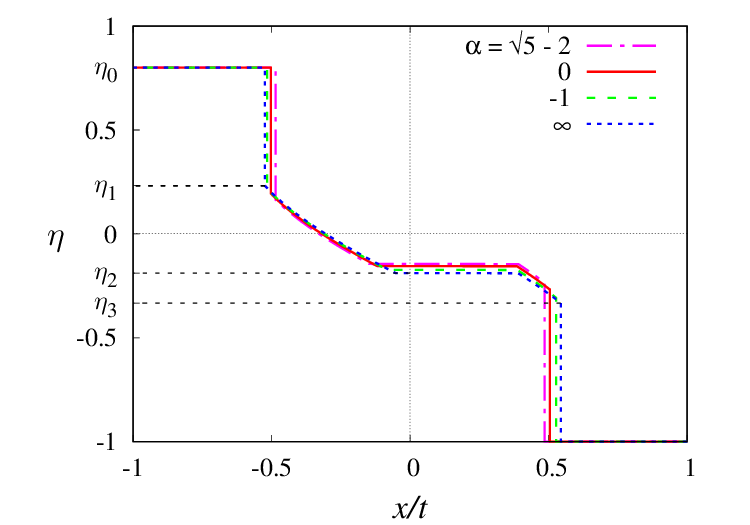}\put(-15,25){(a)}\includegraphics[width=0.5\columnwidth]{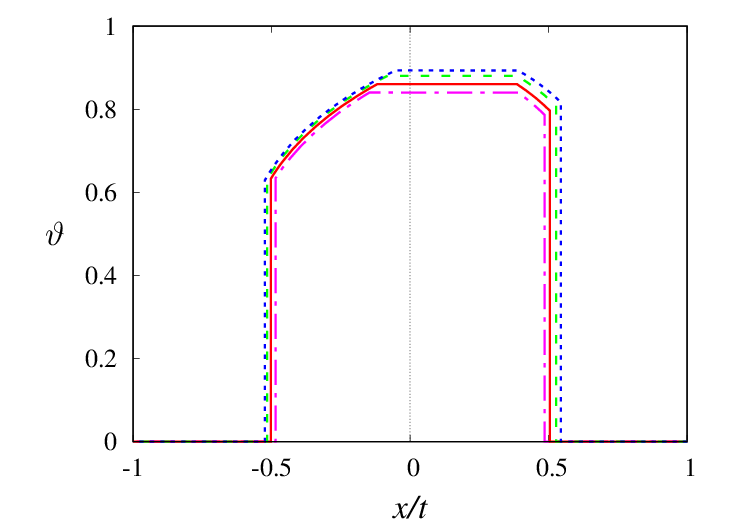}\put(-15,25){(b)}\caption{\label{fig:lckx-mod}The interface height in the modified partial
lock-exchange flow (a) and the corresponding shear velocity (b) for
the lock height $h_{0}=(1+\eta_{0})/2=0.9.$}
\end{figure}

As for the full lock exchange, the upstream and downstream solutions
connect without jump at the point of equal shear velocities $\vartheta_{+}=\vartheta_{-}$
(see Fig. \ref{fig:lckx-mod}). This yields $\theta_{2}=(r^{+}-r^{-})/2$
with the Riemann invariants defined by Eqs. (\ref{eq:r-},\ref{eq:r+}).
The corresponding interface height $h_{2}=(1+\eta_{2})/2,$ where
$\eta_{2}=\sin((r^{+}-r^{-})/2),$ is plotted in Fig. \ref{fig:lckx-mod-h01}(a)
against the lock height $h_{0}.$

\begin{figure}
\centering{}\includegraphics[width=0.5\columnwidth]{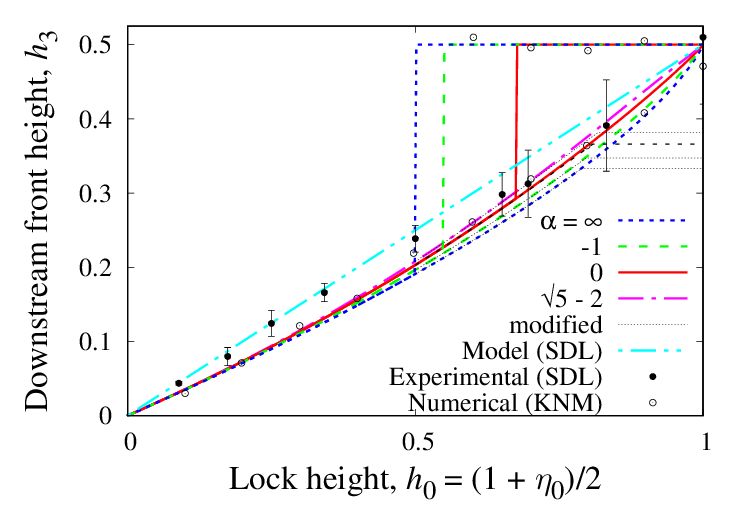}\put(-10,25){(a)}\includegraphics[width=0.5\columnwidth]{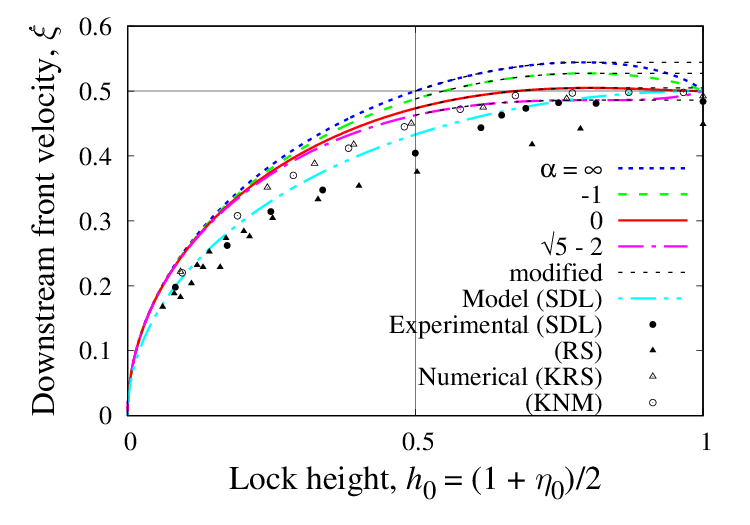}\put(-10,25){(b)}\caption{\label{fig:lckx-numex} The front height (a) and velocity of the downstream
gravity current (b) for various $\alpha$ against the lock height:
comparison of the modified and original lock-exchange solutions with
the experimental results of \citet[(SDL)][]{Shin2004} and \citet[(RS)][]{Rottman1983};
the numerical results of \citet[(KRS)][]{Klemp1994} and \citet[(KNM)][]{Khodkar2017};
the model of \citet[(SDL)][]{Shin2004}. The abrupt variation of the
front height in the original lock-exchange solution is due to the
disappearance of the head block at the critical upstream interface
height (\ref{eq:etac}).}
\end{figure}

The above solution holds only for sufficiently high locks which produce
an intermediate interface height $\eta_{2}$ not lower than the downstream
front height $\eta_{3}$ for a given $\alpha.$ For lower locks, the
upstream solution can connect directly to the quiescent downstream
state. The downstream front heights and velocities resulting from
the jump condition for various $\alpha$ are plotted in Fig. \ref{fig:lckx-numex}
along with the solution of the original lock-exchange problem. This
figure also shows the relevant numerical and experimental results.
As seen, the difference between the modified and original lock-exchange
solutions is significant only for the fronts of super-critical height.
Note that this is the height at which the front speed for a given
$\alpha$ attains maximum. The height of gravity current predicted
by the SW model can be seen to agree well with the DNS results of
\citet{Khodkar2017} while the experimental results of \citet{Shin2004}
for shallow locks $(h_{0}\lesssim0.5)$ are somewhat higher and close
to the half lock height predicted by their theoretical model. It is
also interesting to note that the DNS results of \citet{Khodkar2017}
feature a gravity current head which is present for tall locks $(h_{0}>0.5)$
and extends up to the channel mid-height as predicted by the SW solution.

\section{\label{sec:num}Numerical simulation using conservative SW equations}

In this section, we verify the analytical solution obtained in Sec.
\ref{sec:analytic} by solving the ideal lock-exchange problem numerically
using the SW mass and generalized momentum conservation equations
(\ref{eq:vlm},\ref{eq:gen}) and the LWLF4 composite scheme in which
three steps of Lax-Wendroff scheme are followed by a step of Lax-Friedrichs
scheme. \citep{Liska1998} This composition significantly reduces
numerical oscillations around the jumps, which are typical to the
Lax-Wendroff scheme, without introducing excessive numerical diffusion,
which is typical to the Lax-Friedrichs scheme. The generalized momentum
equation (\ref{eq:gen}) is solved for $\alpha=0,-1,10^{10},$ where
the last value effectively reduces the momentum equation to the circulation
conservation equation (\ref{eq:vrt}) which is formally recovered
in the limit $\alpha\rightarrow\infty.$

\begin{figure}
\begin{centering}
\includegraphics[width=0.5\columnwidth]{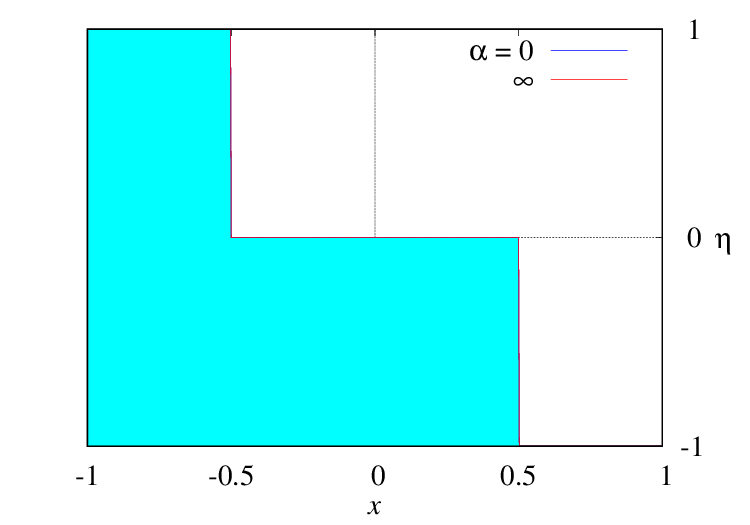}\put(-42,28){(a)}\includegraphics[width=0.5\columnwidth]{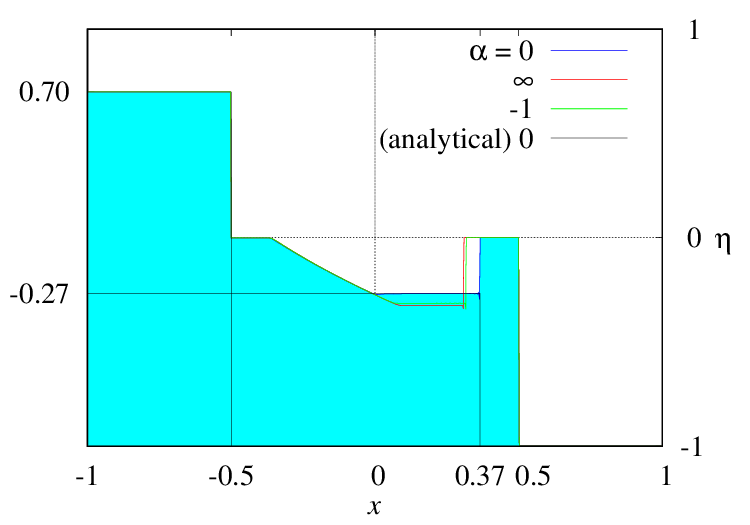}\put(-42,28){(b)}
\par\end{centering}
\centering{}\includegraphics[width=0.5\columnwidth]{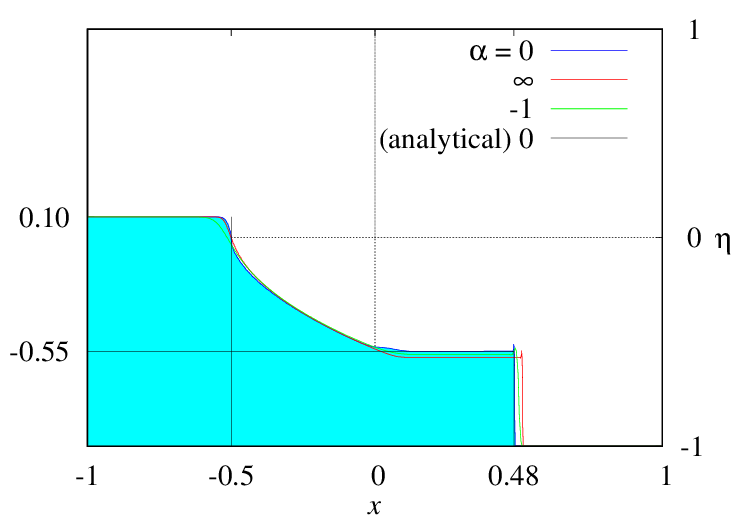}\put(-42,28){(c)}\includegraphics[width=0.5\columnwidth]{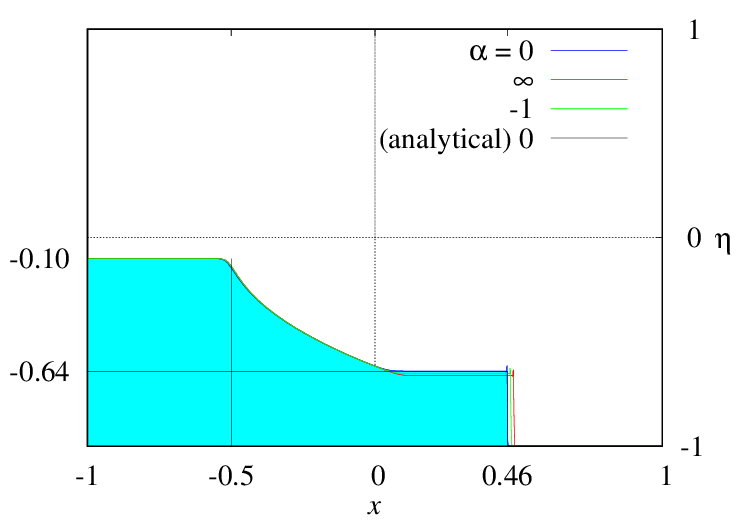}\put(-42,28){(d)}\caption{\label{fig:nmr}The interface height at the time instant $t=1$ after
opening the lock gate of height $\eta_{0}=1$ (a), $0.7$ (b), $0.1$(c)
and $-0.1$(d) computed numerically by solving the SW mass and generalized
momentum conservation equations (\ref{eq:vlm},\ref{eq:gen}) with
$\alpha=0$ (filled area), $-1$ and $\infty$ using the LWLF4 scheme
with the time step $\tau=10^{-3}$ and the spatial step $\delta=10^{-3}$
(a,b) and $\tau=2.5\times10^{-4}$, $\delta=1.25\times10^{-4}$ (c,d).}
\end{figure}

It is important to note that the integration of Eq. (\ref{eq:gen})
is hindered by the product $(\eta+\alpha)\vartheta=w$ which appears
as a dynamical variable. This quantity has to be determined along
with $\eta$ in each time step and then used to calculate the shear
velocity as $\vartheta=w/(\eta+\alpha).$ The division by $\eta+\alpha$
in the last step can produce large numerical errors at the points
where $\eta$ happens to be close to $-\alpha,$ which is possible
when $|\alpha|<1.$ Although this numerical uncertainty can formally
be resolved using the L'Hôpital's rule, it was not possible to construct
a stable numerical scheme in this way. Since there is no such a problem
in the circulation conservation equation (\ref{eq:vrt}) and the latter
is equivalent to the momentum equation at smooth parts of the interface,
the division by zero was avoided by adopting a hybrid approach in
which Eq. (\ref{eq:vrt}) is used instead of Eq. (\ref{eq:gen}) at
the grid points where $|\eta+\alpha|<\varepsilon\approx10^{-3}$ to
find $\vartheta$ directly. This approach was found to produce numerical
results in good agreement with the analytical solution for a range
of lock heights (see Fig. \ref{fig:nmr}).

First, as seen in Fig. \ref{fig:nmr}(a), the exact solution for the
full lock-exchange is reproduced using equal time and space steps.
This is an optimal choice which renders the scheme marginally stable
and ensures that the front advances one grid step in one time step.
In general, such marginally stable schemes are known to reduce spurious
oscillations at the jumps. \citep{Lerat1974} The scheme becomes unstable
at larger time steps which violate the Courant-Friedrichs-Lewy condition,
whereas spurious oscillations arise at smaller time steps. In both
cases, the numerical solution for the full lock-exchange breaks down.

Using the same time step and grid size as for the full lock-exchange,
we were able to reproduce the exact solution also for a range of partial
lock-exchange flows. The numerical solutions for the lock height $\eta_{0}=0.7$
is shown in Fig. \ref{fig:nmr}(b) for $\alpha=0,-1,\infty$ along
with the key parameters of the analytical solution for $\alpha=0.$
In this case, spurious oscillations appear behind the head block because
the trailing jump advances less than one grid step per time step.
As the head block becomes progressively thinner with lowering $\eta_{0},$
there is a range lock heights $0.1\lesssim\eta_{0}\lesssim0.7$ for
which it was not possible to find a numerically stable solution. The
stable numerical solution that re-emerges at $\eta_{0}\approx0.1$
has no head block but just the downstream and upstream jumps. The
latter can be seen in Fig. \ref{fig:nmr}(c) to be somewhat smoothed
out by the numerical diffusion produced by the Lax-Friedrichs step
of the LWLF4 scheme. This upstream jump vanishes when the lock is
lower than channel mid-height $(\eta_{0}\le0).$ The corresponding
numerical solution shown in Fig. \ref{fig:nmr}(c) can be seen to
produce a finite front height as predicted by the analytical solution.
It is important to note that in this case, also a smooth analytical
solution akin to the single-layer dam-break solution is in principle
possible. However, as argued above, such a smooth solution is unstable
and, thus, not observable in the numerical solution.

\section{\label{sec:Sum}Summary and conclusion}

In the present paper, we considered the lock-exchange flow which is
triggered by rapidly removing a vertical gate between two fluids of
slightly different densities contained in a horizontal channel bounded
by a rigid lid. Using the method of characteristics and analytic expressions
for the Riemann invariants of the underlying system of two hyperbolic
differential equations, we obtained a simple-wave solution for this
problem. The multivaluedness as well as the instability of the solution
was found to result in a number of hydraulic jumps which were resolved
using the conservation of mass and impulse (pseudo-momentum). The
respective Rankine-Hugoniot jump conditions contain a free parameter
$\alpha,$ which defines the relative contribution of each layer to
the pressure gradient along the interface in the generalized SW momentum
equation. \citep{Priede2020} We considered the solution for $\alpha=0,$
which corresponds to the interfacial pressure gradient determined
by both layers with equal weight coefficients, along with the solutions
for $\alpha=-1$ and $\alpha\rightarrow\infty$ which reproduce the
classic front condition of \citet{Klemp1997} and the circulation
conservation condition of \citet{Borden2013}.

For the full-height lock exchange, which corresponds to the heavier
fluid behind the lock gate occupying the whole channel, the solution
does not depend on $\alpha$ and consists of a downstream gravity
current and a symmetric upstream bore which span the upper and lower
halves of the channel and propagate at the dimensionless speed (Froude
number) equal to $1/2.$ This solution, which conserves not only the
mass and impulse but also the circulation and energy, coincides with
that assumed by \citet{Yih1955} but differs from the numerical solution
obtained by \citet{Klemp1994} who assume initial state to be a gravity
current rather than a quiescent fluid.

The upstream part of the solution remains independent of $\alpha$
also for partial-height lock exchange. The reduction of the lock height
just reduces the upstream bore height which elevates above the channel
mid-height and keeps propagating upstream at a constant speed -1/2.
This bore, which is fully conservative as for the full lock-exchange,
persists as long as the upstream interface is located above the channel
mid-height. In this case, the downstream gravity current features
an elevated head block which extends up to the channel mid-height.
Both the upstream bore and the elevated downstream head are in a good
agreement with the 2D DNS results of \citet{Khodkar2017}. 

The solution with the head block, however, is possible only for the
upstream interface heights above a certain minimal height which depends
on $\alpha.$ The head block is connected to the interface depression
behind it by a back-step-type jump whose speed of propagation increases
with lowering the upstream interface height until, at a certain critical
height depending on $\alpha,$ it reaches the leading front speed
$1/2$. At this critical point, the head block vanishes and the upstream
solution connects directly to the quiescent downstream state.

Since the upstream state can connect directly to the quiescent downstream
state without the head block also at larger upstream interface heights,
multiple solutions are in principle possible. If the ensuing gravity
current exceeds a certain critical height depending on $\alpha$ at
which its speed of propagation attains maximum, physical considerations
imply that such a gravity current is unstable. \citep{Priede2020}
This instability can result in the increase in the height of gravity
current up to the channel mid-height so giving rise to the elevated
head which is observed also in the 2D DNS results. \citep{Khodkar2017}

Alternatively, this instability may cause the front of a supercritical
gravity current to collapse to a lower height corresponding to the
maximal velocity of propagation for a given $\alpha.$ Such a collapse
could break the dependence of the solution on the initial conditions
and thus to lead to a new initial state which contains a gravity current
of lower depth. This assumption underlies the alternative formulation
of the lock exchange problem which was also considered in this paper.
Although the new front height is not uniquely defined, the problem
can be solved analytically using the method of characteristics. The
resulting gravity current speed can exceed $\frac{1}{2},$ especially
when the conservation of circulation with $\alpha\rightarrow\infty$
or the momentum conservation with $\alpha=-1$ are assumed. This result
is at odds with experimental observations as well as with numerical
simulations which show the front speed to be somewhat lower than $\frac{1}{2}.$
A much better agreement with highly accurate numerical results is
achieved by $\alpha=\sqrt{5}-2,$ which produces the largest stable
front height and yields the front speed $\dot{\xi}=\alpha^{1/2}\approx0.486.$
This value of $\alpha$ can be selected dynamically by the collapse
of unstable gravity current stopping at at the largest possible front
height.

The exact solution, which represents an inviscid approximation to
high-Reynolds-number limit, can be used as a benchmark for SW numerical
schemes. In the present paper, we demonstrated that two-layer SW equations
for the mass and generalized momentum conservation can describe interfacial
gravity waves containing hydraulic jumps in a self-contained way without
invoking external closure conditions. This is a definite achievement
as previously two-layer SW equations were thought to be inherently
incomplete and requiring externally derived front conditions in order
to describe hydraulic jumps.

\section*{Author declarations}

\subsection*{Conflict of Interest}

The authors have no conflicts to disclose.

\bibliography{/home/priede/work/swe2d1l/ref/shwt}

\end{document}